\documentclass[cameraready]{Interspeech}

\title{Full-Duplex Interaction in Spoken Dialogue Systems: A Comprehensive Study from the ICASSP 2026 HumDial Challenge}

\author[affiliation={1}]{Chengyou}{Wang}
\author[affiliation={1}]{Hongfei}{Xue}
\author[affiliation={1}]{Guojian}{Li}
\author[affiliation={1}]{Zhixian}{Zhao}
\author[affiliation={1}]{Shuiyuan}{Wang}
\author[affiliation={2}]{Shuai}{Wang}
\author[affiliation={3}]{Xin}{Xu}
\author[affiliation={3}]{Hui}{Bu}
\author[affiliation={1}, correspondingauthor]{Lei}{Xie}

\address{
    $^1$ Audio, Speech and Language Processing Group (ASLP@NPU), School of Computer Science, Northwestern Polytechnical University, Xi’an, China \\
    $^2$ Nanjing University, China \\
    $^3$ AISHELL, China
}

\email{asd6404112a@mail.nwpu.edu.cn, hfxue@mail.nwpu.edu.cn, lxie@nwpu.edu.cn}


\usepackage{comment}
\usepackage{multirow}
\usepackage[table]{xcolor}

\begin{document}

\maketitle

\begin{abstract}

Full-duplex interaction, where speakers and listeners converse simultaneously, is a key element of human communication often missing from traditional spoken dialogue systems. These systems, based on rigid turn-taking paradigms, struggle to respond naturally in dynamic conversations. The Full-Duplex Interaction Track of ICASSP 2026 Human-like Spoken Dialogue Systems Challenge (HumDial Challenge) aims to advance the evaluation of full-duplex systems by offering a framework for handling real-time interruptions, speech overlap, and dynamic turn negotiation.
We introduce a comprehensive benchmark for full-duplex spoken dialogue systems, built from the HumDial Challenge. We release a high-quality dual-channel dataset of real human-recorded conversations, capturing interruptions, overlapping speech, and feedback mechanisms. This dataset forms the basis for the HumDial-FDBench benchmark, which assesses a system's ability to handle interruptions while maintaining conversational flow.
Additionally, we create a public leaderboard to compare the performance of open-source and proprietary models, promoting transparent, reproducible evaluation. These resources support the development of more responsive, adaptive, and human-like dialogue systems.

\end{abstract}

\section{Introduction}
In natural communication, human dialogue rarely follows strict turn alternation. Instead, speakers and listeners interact through a continuous full-duplex process in which listening and speaking may occur simultaneously. Participants regulate conversational flow through cues such as intonation, pauses, prosody, and semantic context. These signals allow interlocutors to interrupt, confirm, elaborate, or revise an utterance while simultaneously providing feedback through brief acknowledgments or backchannels. As a result, conversational control emerges dynamically rather than through a predetermined sequence of turns.

By contrast, many existing speech-based dialogue systems~\cite{leviathan2018google,wang2024freeze,fu2025vita,defossez2024moshi,geng2025osum} still rely on rigid turn-taking paradigms. System responses are typically generated only after explicit detection of speech completion or based on simple voice activity signals used to infer turn boundaries. Such sequential interaction disrupts conversational continuity and limits system responsiveness when handling unexpected inputs, redundant instructions, or contextual changes. Consequently, system behavior in complex real-world environments often lacks the flexibility and responsiveness required for natural human-like interaction. Although full-duplex interaction has recently attracted increasing attention in both academia and industry, research progress remains constrained by limitations in available datasets and evaluation methodologies.

From the perspective of data resources, most public dialogue corpora consist of single-channel recordings or task-oriented scripted interactions that do not fully capture realistic conversational dynamics. Phenomena such as overlapping speech, speaker interruption, background interference, and contextual shifts following interruptions frequently occur in real conversations but are often simplified or omitted in existing datasets. In addition, paralinguistic cues such as emotional variation, backchannels, and pause patterns are commonly annotated with limited granularity, restricting the ability of models to learn realistic turn negotiation behavior. From the perspective of evaluation, there is still no widely adopted benchmark for assessing key capabilities of full-duplex dialogue systems. Existing metrics typically focus on recognition accuracy or task completion rates, overlooking interactional dimensions such as interruption handling, response timing, and the ability to resume dialogue after disruption. This misalignment between evaluation criteria and interaction behavior limits fair comparison across systems and slows progress toward deployable full-duplex dialogue models.


The Full-Duplex Interaction Track of the ICASSP 2026 Human-like Spoken Dialogue Systems Challenge~\cite{DBLP:journals/corr/abs-2601-05564} aims to advance the evaluation of full-duplex dialogue systems by introducing a dual-channel dialogue dataset~\footnote{\url{https://github.com/ASLP-lab/HumDial-FDBench}} of real human-recorded conversations. The dataset captures realistic conversational phenomena such as interruptions, overlapping speech, and dynamic turn negotiation. Based on this dataset, we establish HumDial-FDBench, a benchmark designed to evaluate a system's ability to handle interruptions and maintain conversational continuity during concurrent listening and generation.
In addition to releasing the dataset as an open resource, we build a public leaderboard that reports benchmark results from both open-source and proprietary models, as well as submissions from participating teams in the challenge. This paper presents the dataset, evaluation methodology, and baseline systems used in the track, and summarizes the approaches explored by participating teams. Together, these resources provide a shared benchmark for studying full-duplex spoken dialogue interaction and support future research toward more responsive and human-like conversational systems.
\section{Related Work}

\textbf{Full-Duplex Spoken Interaction.} Mainstream spoken dialogue system~\cite{yu2024salmonn,defossez2024moshi} architectures can be broadly categorized into cascaded pipelines and end-to-end frameworks. Cascaded systems typically chain modules such as ASR, LLM, and TTS to complete the interaction, as exemplified by Google-Duplex~\cite {leviathan2018google} and Firered-Chat~\cite{chen2025fireredchat}. However, this design often incurs high end-to-end latency and suffers from progressive error accumulation across modules. In recent years, end-to-end models emerge as a more streamlined alternative, including Freezi-Omni~\cite{wang2024freeze}, Moshi~\cite{defossez2024moshi}, and Osum-EChat~\cite{geng2025osum}. Nevertheless, most end-to-end systems still rely on additional interruption-related modules or strategies for turn decision, such as voice activity detection (VAD), TEN-Turn~\footnote{\url{https://github.com/TEN-framework/ten-turn-detection}}, and Smart-Turn~\footnote{\url{https://github.com/pipecat-ai/Smart-Turn}}, and a unified and stable modeling paradigm for full-duplex interaction remains absent.

\noindent \textbf{Benchmarks for Dialogue Systems.} Existing benchmarks for dialogue systems often struggle to meet the authenticity requirements of real-world interaction and also provide limited fine-grained quantitative measures for complex behaviors, such as subtle rejection behaviors or feedback signals. Early benchmarks~\cite{du2025mtalk} mainly focus on single-turn or half-duplex settings, which makes it difficult to characterize the dynamic interaction process associated with speech overlap. Full-duplex bench~\cite{lin2025full} is the first to introduce an evaluation perspective centered on overlap handling, whereas FD-bench~\cite{peng2025fd} further extends the scope to more complex interaction scenarios and longer multi-turn settings, bringing the evaluation of full-duplex systems closer to practical dialogue demands.

\section{Released Dataset}
\subsection{Scenarios}
As shown in Table~\ref{tab:scenario_stats}, the released dataset includes two major categories: Interruption and Rejection, with a total of eight scenarios.

\textbf{Interruption}: The Interruption scenario assesses the capability of a model to adapt an ongoing response when the user intervenes. It includes five sub-scenarios. \textit{Follow-up Questions} cover cases where the user interrupts to ask a related question and expects an immediate and relevant response. \textit{Negation or Dissatisfaction} refers to situations where the user expresses disagreement or dissatisfaction mid-utterance, which requires a prompt adjustment or correction of the response. \textit{Repetition Requests} describe interruptions caused by inaudibility or misunderstanding, where the user asks the model to repeat the previous output. \textit{Topic Switching} considers abrupt shifts to a new topic, which requires a smooth and coherent transition in the response. \textit{Silence or Termination} captures cases where the user explicitly asks the model to stop speaking, and the model is expected to immediately cease output while remaining ready to resume upon further prompting.

\textbf{Rejection}: The Rejection scenario focuses on the capability of a model to withhold responses to non-actionable or irrelevant speech. It includes four sub-scenarios. \textit{User Real-time Backchannels} involve short acknowledgments (e.g., ``uh-huh,'' ``yeah'') that should not interrupt an ongoing response. \textit{Pause Handling} covers hesitations or pauses mid-sentence, where the model is expected to wait until the user's intent is fully expressed before responding. \textit{Third-party Speech} refers to background speakers who interject before or after the user query, and the model is expected to reject these utterances and respond only to the target user. \textit{Speech Directed at Others} captures cases where the user temporarily addresses another person, often on an unrelated topic, and the model is expected to detect and ignore such utterances.

\begin{table}[ht]
  \caption{Statistics of the dataset. 
  Numbers denote instances per split.}
  \label{tab:scenario_stats}
  \centering
  \scriptsize
  \begin{tabular}{p{0.12\columnwidth}p{0.37\columnwidth}ccc}
    \toprule
    Category & Scenario & Train & Dev & Test \\
    \midrule
    Interruption
    & Follow-up Question     & 1507 & 200 & 600 \\
    & Negation or Dissatisfaction & 1211 & 200 & 600 \\
    & Repetition Request     & 1213 & 200 & 600 \\
    & Topic Switch           & 1213 & 200 & 600 \\
    & Silence or Stop        & 1212 & 200 & 600 \\
    \midrule
    Rejection
    & User Real-time Backchannels            & 1211 & 200 & 600 \\
    & Pause Handling         & 1211 & 200 & 600 \\
    & Third-party Speech     & 120  & 200 & 600 \\
    & Speech Directed to Others & 0 & 200 & 200 \\
    \bottomrule
  \end{tabular}
\end{table}

\subsection{Data construction}
We adopt a two-stage construction strategy that combines LLM script generation with human recording to ensure authenticity and controllability in interaction behavior, semantic coherence, and acoustic naturalness. Specifically, DeepSeek~\cite{liu2024deepseek} generates natural dialogue scripts that embed targeted interaction cues, including intrusive interjections, aside-style responses, and brief utterances directed to another person, which ensures coverage of key full-duplex phenomena at the script level. Professional actors then perform and record these scripts to reproduce the dynamics of real conversation, where concurrent listening and speaking, barging-in, hesitation pauses, and feedback signals co-occur. In contrast to synthetic mixing that overlaps audio tracks on a timeline, this procedure preserves natural overlap timing, prosodic variation, and interaction rhythm, thereby providing a data foundation that is closer to real-world settings for full-duplex modeling and evaluation.

Based on this pipeline, we release a real-world speech dataset for full-duplex dialogue systems. The dataset contains more than 100 hours of human-recorded interactive speech and captures realistic conversational dynamics. It provides a detailed annotation scheme and complete train--validation--test splits. The dataset covers both Chinese and English, which supports cross-lingual studies and enables standardized and reliable benchmarking for quantitative evaluation of full-duplex systems.

\section{HumDial-FDBench}

The HumDial-FDBench evaluation protocol is built upon Full-Duplex-Bench v1.5~\cite{lin2025full1.5} and introduces several extensions to support more complex interaction scenarios and a more comprehensive assessment of full-duplex dialogue systems.

\subsection{Behavior evaluation}

To analyze the semantic response strategy of a model following speech overlap, we obtain time-aligned ASR transcripts using Paraformer~\cite{gao2022paraformer} for Chinese and Parakeet-TDT for English. Based on these transcripts, DeepSeek-V3~\cite{liu2024deepseek} is employed to categorize the model responses using carefully designed prompts. Following the protocol of Full-Duplex-Bench v1.5, responses are classified into four categories: \textit{Respond}, \textit{Resume}, \textit{Uncertain}, and \textit{Unknown}.

\subsection{Evaluation metrics}

We design evaluation metrics to assess system performance under different conversational scenarios, including interruption handling, rejection behavior, and response latency.

\textbf{Interruption.} In interruption scenarios, only \textit{Respond} is considered correct, as successful interruption requires promptly addressing the user's intent within the overlapping speech, ensuring real-time response to the user's immediate needs. 

\textbf{Rejection.} For backchannels, third-party speech after the user's question, and speech where the user turns to address another person, we follow the interruption protocol, but only \textit{Resume} is correct. Successful rejection requires ignoring invalid inputs while maintaining the dialogue state. For Pause handling and third-party speech before the user's question, we use a customized criterion based on word-level timestamps from the ASR transcripts. If the model responds before the user's semantic content is complete, or in third-party-first scenarios where the model responds before the user’s question, the rejection is considered unsuccessful. This process results in a binary rejection outcome, used to compute overall rejection accuracy.
\begin{figure}[htbp]
    \centering
    \includegraphics[width=\columnwidth]{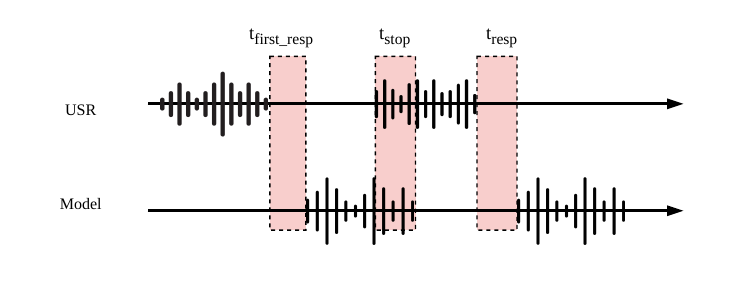} 
    \caption{Framework for latency evaluation in interruption scenarios.}
    \label{fig:framework}
\end{figure}

\textbf{Latency Evaluation.} In addition to evaluating behavioral correctness, we measure real-time responsiveness in interruption scenarios, as shown in \autoref{fig:framework}. To enhance our evaluation, we introduce a new metric, first response latency, alongside stop latency and response latency. First response latency is defined as the interval from the end of the first user question to the start of the first model response, which helps assess the model's responsiveness in real-time turn-taking scenarios.
    \begin{equation}
    t_{\text{first\_resp}} = t_{\text{first\_model\_start}} - t_{\text{first\_user\_end}} .
    \end{equation}

All latency metrics ($t_{\text{stop}}$, $t_{\text{resp}}$, and $t_{\text{first\_resp}}$) are reported in seconds. Speech activity boundaries are detected consistently using Silero-VAD~\cite{SileroVAD}.
In rejection scenarios, only the first response latency is measured. This metric is computed for successful rejections, reflecting the response speed to the true user intent, and is defined as the interval from the end of the true user question to the start of the first valid model response.

\subsection{Overall Score}  

The latency score evaluates system responsiveness, particularly in real-time scenarios such as interruption handling. It is computed as the mean of the stop latency, response latency, and first response latency across their respective sub-scenarios.
This average value is denoted as \( L \). To ensure comparability with the behavioral scores, logarithmic normalization is applied:

In this formula, \( L_{\text{min}} \) represents the minimum latency observed across all systems, and \( L_{\text{base}} \) corresponds to the baseline system’s latency, set to 60. This normalization aligns the latency score with the behavioral scores.

The interruption and rejection scores are calculated as the arithmetic mean of the behavioral scores across their respective sub-scenarios. These scores are given higher weight due to their importance in real-time dialogue management, while the latency score contributes less to the overall score.

The final score, \( S_{Total} \), is a weighted combination of the interruption score \( S_{Int} \), the rejection score \( S_{Rej} \), and the latency score \( S_{Delay} \):
\begin{equation}
S_{Delay}(L) = 100 - 40 \times \frac{\log \left( \frac{L}{L_{\text{min}}} \right)}{\log \left( \frac{L_{\text{base}}}{L_{\text{min}}} \right)}.
\end{equation}

\begin{equation}
S_{Total} = 0.4 \times S_{Int} + 0.4 \times S_{Rej} + 0.2 \times S_{Delay}.
\end{equation}

\section{Leaderboard}

\subsection{Model Evaluation}

We evaluated a diverse set of dialogue management models, divided into two categories: Open-Source Models, such as Freeze-Omni~\cite{wang2024freeze}, Moshi~\cite{defossez2024moshi}, and Osum-EChat, which provide transparency and reproducibility, and Closed-Source Models, like Gemini 2.5~\cite{comanici2025gemini}.
To ensure consistent evaluation across systems, we use a baseline that integrates Easy-Turn~\cite{DBLP:journals/corr/abs-2601-20230} with Osum-EChat~\cite{geng2025osum}.

\subsection{Experimental Results}

\begin{table}[htb]
  \caption{Results on HumDial-FDBench of different models. \textit{D-Sco.} stands for Delay Score, representing the system's performance in handling delay-related metrics. 
  * indicates late submission. \colorbox{green!20}{green} rows open-source models,\colorbox{gray!20}{gray} rows closed-source models}
  \label{tab:track2_results}
  \centering
  \small
  \setlength{\tabcolsep}{3pt}
  \begin{tabular}{p{2cm}ccccc c}
    \toprule
    Team & Int. & Rej. & Delay (s) & D-Sco. & Final & Rank \\
    \midrule
    Cookie asr      & 79.3 & 72.2 & 1.260 & 79.9 & 76.6 & 1 \\
    Badcat~\cite{DBLP:journals/corr/abs-2601-20230}          & 89.7 & 57.8 & 1.632 & 72.6 & 73.5 & 2 \\
    SenseDialog     & 76.4 & 60.9 & 1.237 & 80.5 & 71.0 & 3 \\
    \rowcolor{gray!20}Gemini-2.5       & 79.8 & 36.5 & 1.301 & 79.0 & 62.3 & --\\
    Unity Squad*    & 68.5 & 51.2 & 1.876 & 68.6 & 61.6 & -- \\
    RhythmSense     & 77.4 & 38.6 & 1.577 & 73.5 & 61.1 & 4 \\
    Lingcon Insight & 67.6 & 38.9 & 1.127 & 83.1 & 59.2 & 5 \\
    Baseline        & 75.9 & 35.2 & 2.531 & 60.0 & 56.4 & 6 \\
    HelloWorld      & 51.3 & 36.3 & 0.624 & 100.0 & 55.0 & 7 \\
    \rowcolor{green!20}Freeze-Omni  & 29.6 & 50.2 & 2.578 & 59.5 & 43.8 & --\\
    AISpeech        & 47.7 & 33.9 & 3.391 & 51.6 & 43.0 & 8 \\
    Cascade         & 28.1 & 30.9 & 1.739 & 70.7 & 37.7 & 9 \\
    \rowcolor{green!20}Moshi        & 35.4 & 22.8 & 2.876 & 56.3 & 34.5 & -- \\

    \bottomrule
  \end{tabular}
\end{table}
Table~\ref{tab:track2_results} shows that Gemini outperforms open-source models in interruption handling with a score of 79.8, while Moshi and Freeze-Omni score lower (35.4 and 29.6, respectively). In rejection handling, Freeze-Omni leads with a score of 50.2, while Moshi and Gemini score lower (22.8 and 36.5). Gemini also has the lowest latency (1.301 seconds) due to its VAD-based interruption strategy.

Despite progress in performance, systems still face structural challenges, particularly in controlling response latency under complex or uncertain acoustic conditions. Delays in turn decisions disrupt conversational timing and fluency. Performance also degrades in multi-speaker environments, where overlapping speech from third parties complicates speaker identification and leads to incorrect decisions. Additionally, robustness under background noise remains limited, with transient noises causing false activations or missed detections.

Achieving real-time, fluent full-duplex interaction requires balancing latency control with stability under challenging conditions. While streaming interruptions can reduce delays and improve responsiveness, maintaining decision accuracy with continuous input and incomplete semantic evidence remains a challenge. Real-world scenarios with speaker overlap and background noise require better speaker discrimination and noise suppression to develop stable, natural full-duplex systems.

\section{Analysis}

This section summarizes the representative design choices of Track~II submissions from three complementary perspectives: architecture paradigm, turn-taking strategy, and training strategy. The goal is to highlight the major design axes and the practical trade-offs that shaped system performance under full-duplex interaction.

\subsection{System Architecture}
The submissions in this challenge cover three main architecture paradigms widely used in dialogue systems: cascaded, semi-cascaded with unified decision-making, and end-to-end (E2E) multimodal systems.

\textbf{Cascaded}. Most submissions adopt cascaded or semi-cascaded pipelines. In this design, turn-taking decisions are made by a lightweight front-end module and processed by a response-generation back-end. A typical pipeline includes VAD-based speech segmentation, streaming or partial ASR for lexical cues, a decision component for interruption or rejection, and a TTS module for speaking. This architecture remains dominant due to its modularity and controllability, especially under streaming constraints. Teams such as \textbf{cookie\_asr}, \textit{Unity Squad}, \textit{RhythmSense}, and \textit{Cascade} use this approach.

\textbf{Semi-Cascaded}. Some systems use semi-cascaded pipelines with unified decision-making. These systems combine an independent perception module and a unified decision-making "brain" for control. Teams like \textit{Badcat} and \textit{SenseTime} have adopted this approach. While balancing flexibility and end-to-end capability, it tends to increase system complexity.

\textbf{End-to-End}. The E2E multimodal LLM~\cite{yu2024salmonn,xu2025qwen3,zeng2024glm} architecture, combining both perception and generation in a unified model, is used by teams like \textit{Lingcon Insight} and \textit{HelloWorld}. This approach offers low latency and the ability to perceive prosodic cues, but faces challenges like "auditory blindness" during generation and difficulty in controlling the output. E2E models require careful design to balance performance and reliability.

\subsection{Turn-Taking Strategy}

Turn-taking strategy is the most distinctive factor across submissions, reflecting the core challenge of Track~II: distinguishing between interruption, backchannel (e.g., short acknowledgements), and rejection cases such as noise, side speech, or third-party utterances.

\textbf{Heuristic Rules.} This strategy relies on handcrafted logic, such as lexical triggers, semantic similarity checks, and time-duration thresholds. \textit{RhythmSense} adopts a multi-stage arbitration scheme that first checks lexical cues, then falls back to semantic matching, and finally applies conservative rules when evidence is weak. These methods require little to no training and can be competitive under certain dataset distributions, but they often struggle in complex acoustic conditions.

\textbf{Specialized Models.} These systems use dedicated lightweight models for turn-related categorization. Representative approaches include a Whisper-tiny-based~\cite{radford2023robust} four-class classifier, TinyBERT~\cite{jiao2020tinybert} text classification combined with rules, and a neural classifier built on discrete audio representations. Such designs are typically robust, as specialized modules filter most irrelevant noise before invoking the LLM, thereby reducing its decision burden.

\textbf{LLM as A Judge.} In this strategy, turn decisions are delegated to a multimodal LLM, which outputs control labels or decision tokens alongside the dialogue text. Representative examples include \textit{Unity Squad}, which uses Qwen2.5~\cite{DBLP:journals/corr/abs-2412-15115} to output both text and classification labels (Complete/Incomplete), and \textit{Badcat}, which uses Qwen3-Omni~\cite{xu2025qwen3} to emit decision tokens (Switch/Continue). This approach benefits from the semantic understanding of LLMs, especially for ambiguous pauses with incomplete semantics, but it can be less sensitive to purely acoustic cues (e.g., third-party speech without strong lexical evidence) than dedicated acoustic classifiers.

\subsection{Training Strategy}

Training strategies vary significantly across teams, influenced by resource constraints and iteration speed. We identify two primary approaches.

\textbf{Extensive Training.} This approach involves full-parameter fine-tuning of multimodal models or adapters, along with training specialized classifiers using large-scale synthesized data. It is effective for optimizing task-specific metrics, especially in improving rejection accuracy with diverse backchannel and third-party patterns.

\textbf{Frozen Pipelines.} Many teams avoid fine-tuning large models, relying instead on prompt engineering, modular design, or external decision logic. This reflects the strong zero-shot capabilities of current open-source multimodal backbones, enabling the development of full-duplex systems with minimal training, provided that perception modules and turn-control logic are well-designed.

\section{Conclusion}

We conduct a comprehensive study based on the Full-Duplex Interaction Track of the ICASSP 2026 Human-like Spoken Dialogue Systems Challenge. We release a dual-channel dataset of real human-recorded conversations to advance research on real-time, full-duplex systems. To facilitate systematic evaluation, we introduce the HumDial-FDBench benchmark, which is designed to assess a system's ability to manage interruptions and maintain conversational continuity during concurrent listening and generation. Additionally, we establish a public leaderboard to enable transparent and reproducible comparisons of both open-source and proprietary models, as well as submissions from the challenge participants. These resources will significantly contribute to the ongoing development of more responsive and human-like dialogue systems.

\section{Generative AI Use Disclosure}

In this study, generative AI tools were solely used for language refinement and editorial support, aimed at improving the clarity, readability, and overall flow of the manuscript. These tools were not involved in the development of the methodology, execution of experiments, generation of results, or formulation of conclusions. All intellectual contributions, including the formulation of core ideas, development of theoretical frameworks, design of methodologies, implementation of experiments, result analysis, and drawing of conclusions, were independently made by the authors without the involvement of generative AI tools. The authors take full responsibility for the intellectual content and scientific integrity of the manuscript. No generative AI tools have been credited as co-authors, and all authors have reviewed and approved the final version of the manuscript.

\bibliographystyle{IEEEtran}
\bibliography{mybib}

\end{document}